\documentclass[twocolumn,letter]{jpsj3}
\usepackage{graphicx}
\usepackage{dcolumn}
\usepackage{bm}
\usepackage{ulem}
\usepackage{braket}
\usepackage{amsmath}
\usepackage{color}

\title{Electric Ferro-Axial Moment as Nanometric Rotator \\ and Source of Longitudinal Spin Current}

\author{Satoru Hayami$^1$, Rikuto Oiwa$^2$, and Hiroaki Kusunose$^2$}
\inst{
$^1$Graduate School of Science, Hokkaido University, Sapporo 060-0810, Japan \\
$^2$Department of Physics, Meiji University, Kawasaki 214-8571, Japan
}

\abst{
An electric ferro-axial moment, which is characterized by a nonzero expectation value of a time-reversal-even axial vector, exhibits distinct spatial-inversion and time-reversal properties from conventional ferroelectric, ferromagnetic, and ferro-magnetoelectric orders.
Nevertheless, physical properties characteristic of the electric ferro-axial moment have been obscure owing to the absence of its conjugate electromagnetic fields.
We theoretically investigate consequences of the presence of the ferro-axial moment on the basis of the symmetry and microscopic model analyses.
We show that atomic-scale electric toroidal multipoles are the heart of the ferro-axial moment, which act as a nanometric rotator against external stimuli.
Furthermore, we propose an intrinsic generation of a spin current parallel to an applied electric field in both metals and insulators.
Our results not only provide a deep microscopic understanding of the role of the ferro-axial moment but also stimulate a new development for functional materials with use of the electric toroidal moment.
}

\begin{document}
\maketitle

{\it Introduction.---}
Symmetry is one of the most important factors to determine physical properties.
The lowering of the symmetry by external stimuli and/or spontaneous phase transitions brings about the functionality in materials.
The breaking of spatial inversion $\mathcal{P}$ (time-reversal $\mathcal{T}$) symmetry leads to a ferroelectric (ferromagnetic) property with an electric polarization (magnetization), while both breakings lead to ferro-magnetoelectric (magnetic toroidal) property to exhibit a linear magneto-electric effect~\cite{Spaldin_0953-8984-20-43-434203,KhomskiiPhysics.2.20,kopaev2009toroidal,Hayami_PhysRevB.90.024432}.
These ferroic systems are microscopically characterized by the uniformly aligned dipole moments with the distinct $\mathcal{P}$ and $\mathcal{T}$ parities:
the electric dipole (time-reversal-even polar vector) $\bm{Q}$ with $(\mathcal{P}, \mathcal{T})=(-1,+1)$, the magnetic dipole (time-reversal-odd axial vector) $\bm{M}$ with $(\mathcal{P}, \mathcal{T})=(+1,-1)$, and the magnetic toroidal (MT) dipole (time-reversal-odd polar vector) $\bm{T}$ with $(\mathcal{P}, \mathcal{T})=(-1,-1)$.
The conditions for active $\bm{Q}$, $\bm{M}$, and $\bm{T}$ in solids are presented on the basis of group theory~\cite{aizu1966possible,aizu1969possible,aizu1970possible,Schmid_doi:10.1080/00150199408245120,wadhawan2000introduction,schmid2008some,litvin2013magnetic,Hayami_PhysRevB.98.165110,cheong2018broken,Watanabe_PhysRevB.98.245129,Yatsushiro_PhysRevB.104.054412}.
Such a symmetry classification has unveiled potentially ferroic materials even without the net dipole moments, such as the (non)collinear antiferromagnet Mn$_3$Sn, $\alpha$-Mn, and Co$X_3$S$_6$ ($X=$ Nb and Ta) showing the giant anomalous Hall effect~\cite{parkin1983magnetic,nakatsuji2015large,Suzuki_PhysRevB.95.094406,ghimire2018large,Tenasini_PhysRevResearch.2.023051,Chen_PhysRevB.101.104418,vsmejkal2020crystal,Akiba_PhysRevResearch.2.043090,Hayami_PhysRevB.103.L180407,Shao_PhysRevApplied.15.024057,Mangelsen_PhysRevB.103.184408,park2021first}.

\begin{figure}[htb!]
\begin{center}
\includegraphics[width=1.0 \hsize]{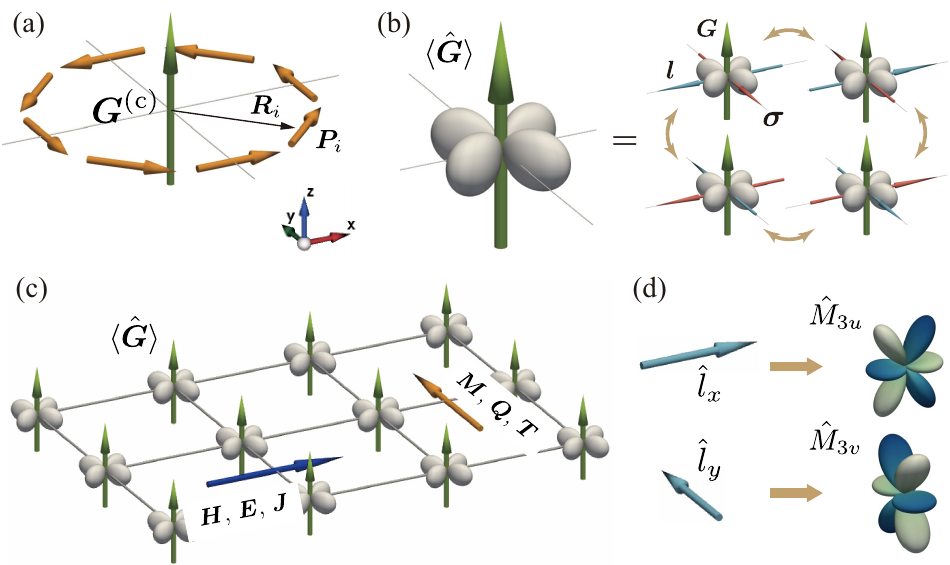}
\caption{
\label{Fig:ETD}
(a) Classical representation of the cluster ET dipole $\bm{G}^{\rm (c)}\leftrightarrow \sum_{i}\bm{R}_{i}\times \bm{P}_{i}$ [see Eq.~(\ref{eq:ETD_classical})].
(b) Quantum-mechanical representation of atomic-scale ET dipole moment $\braket{\hat{\bm{G}}}\leftrightarrow \braket{\hat{\bm{l}}\times \hat{\bm{\sigma}}}$ [see Eq.~(\ref{eq:ETD_quantum})].
The shape of the orbital schematically represents the absence of the vertical mirror symmetry.
(c) Transverse response in the presence of the ET moment; the input external fields ($\bm{H}, \bm{E}, \bm{J}$) induce conjugate physical quantities $(\bm{M}, \bm{Q}, \bm{T})$ along the perpendicular direction.
(d) Correspondence between the magnetic dipole $(\hat{l}_x, \hat{l}_y)$ and octupole $(\hat{M}_{3u}, \hat{M}_{3v})$ belonging to the same irreducible representation.
}
\end{center}
\end{figure}

There is yet another dipole moment, which is so-called the (axial) electric toroidal (ET) dipole characterized by the time-reversal-even axial vector with $(\mathcal{P}, \mathcal{T})=(+1,+1)$~\cite{dubovik1986axial,dubovik1990toroid,Hlinka_PhysRevLett.113.165502}.
In this Letter, we theoretically elucidate that the ET dipole becomes the microscopic origin of an electric ferro-axial (rotational) order, which provides rich physics in contrast to the conventional wisdom.
Although the ET dipole is one of the fundamental dipoles as well as the electric, magnetic, and MT dipoles, its nature has been less investigated for the following two reasons.
One reason is that its quantum-mechanical expression was unknown.
Namely, the ET dipole in solids has originally been introduced as a classical vortex-type object derived from the multipole expansion of the \textit{electric} vector potential~\cite{dubovik1975multipole}:
\begin{align}
\label{eq:ETD_classical}
\bm{G}^{\rm (c)}=c_1\sum_i\bm{R}_i\times \bm{P}_i,
\end{align}
where $\bm{P}_i$ is the electric dipole moment located at a local site $\bm{R}_i$ ($c_1$ is the appropriate coefficient).
The schematic picture of $\bm{G}^{\rm (c)}$ is shown in Fig.~\ref{Fig:ETD}(a).
Although Eq.~(\ref{eq:ETD_classical}) has been widely used to characterize the ferro-axial moment (ordering) in materials~\cite{Johnson_PhysRevLett.108.067201,yadav2016observation,cheong2018broken,jin2020observation,hayashida2020visualization,hanate2021first}, it is applicable only to the cluster system where $\bm{P}_{i}$ and $\bm{R}_{i}$ are independent classical vectors.
Its quantum-mechanical operator expression was discovered quite recently [see Eq.~(\ref{eq:ETD_quantum})], which is then applicable to the atomic-scale wave function~\cite{hayami2018microscopic,kusunose2020complete}.
The other reason is the absence of its conjugate electromagnetic fields owing to the preserved $\mathcal{P}$ and $\mathcal{T}$ symmetries, which naively indicates that the materials with the ET dipole moment would show no intriguing physical responses.

Based on the symmetry and microscopic model analyses, we show that the ET multipoles are active within the atomic $d$-orbital systems.
We find that the nonzero ET moment (an expectation value of the ET multipole) acts as a nanometric rotator against external stimuli;
an input field induces the transverse response of the conjugate physical quantities.
Among such responses, we propose an intrinsic generation of a \textit{longitudinal} spin current by an external electric field.
It is expected to be ubiquitously found since 13 point groups (43 magnetic point groups) possess the axial ET dipole moment~\cite{Hlinka_PhysRevLett.116.177602,Hayami_PhysRevB.98.165110,Yatsushiro_PhysRevB.104.054412,comment_PG}.

\begin{table}[tb!]
\caption{
Four types of dipoles ($\bm{X}=\bm{Q}, \bm{M}, \bm{T}, \bm{G}$) with distinct parities under the spatial inversion ($\mathcal{P}$) and time-reversal ($\mathcal{T}$) operations.
$\bm{X}(\bm{r})$ and $\bm{X}(\bm{k})$ represent the expressions in real and momentum spaces, respectively; $\bm{m}=\bm{l}$ or $\bm{\sigma}$.
The last column represents an effective coupling when $\braket{\hat{\bm{G}}}$ becomes nonzero.
}
\label{tab_irrep}
\centering
\scalebox{0.8}{
\begin{tabular}{ccccccc}
\hline\hline
dipole & symbol & $\mathcal{P}$ & $\mathcal{T}$  & $\bm{X}(\bm{r})$ & $\bm{X}(\bm{k})$ & effective coupling
  \\
\hline
electric & $\bm{Q}$& $-1$ & $+1$ &  $\bm{r}$ &  $\bm{k} \times \bm{\sigma}$ & $\bm{Q}(\bm{r})\times \bm{Q}(\bm{k})$ \\
magnetic & $\bm{M}$ & $+1$ & $-1$ & $\bm{m}$ & $\bm{\sigma}$ & $\bm{M}(\bm{r})\times \bm{M}(\bm{k})$ \\
MT & $\bm{T}$ & $-1$ & $-1$ &  $\bm{r} \times \bm{m}$ & $\bm{k}$ & $\bm{T}(\bm{r})\times \bm{T}(\bm{k})$\\
ET & $\bm{G}$ & $+1$ & $+1$ & $\bm{l}\times \bm{\sigma}$ & --- & ---  \\
\hline \hline
\end{tabular}
}
\end{table}

{\it Axial electric toroidal moment.---}
Let us first discuss the microscopic expression of the ET dipole that does not appear in the conventional multipole expansion.
Equation~(\ref{eq:ETD_classical}) is no longer valid because $\bm{P}_i \parallel \bm{R}_i$ within the atomic orbital; $\bm{G}^{\rm (c)}$ vanishes.
Instead, the atomic-scale ET dipole operator is defined by using two distinct time-reversal-odd axial-vector operators, namely, orbital ($\hat{\bm{l}}$) and spin ($\hat{\bm{s}}=\hat{\bm{\sigma}}/2$) angular momentum operators, which is represented by
\begin{align}
\label{eq:ETD_quantum}
\hat{\bm{G}}=\hat{\bm{l}} \times \hat{\bm{\sigma}}.
\end{align}
This expression was derived by considering the complete basis set to span any of the electronic Hilbert space that describes any atomic-scale electronic degrees of freedom ~\cite{Hayami_PhysRevB.98.165110,kusunose2020complete}.
Equation~(\ref{eq:ETD_quantum}) indicates that the ET dipole moment $\bm{G} \equiv \langle \hat{\bm{G}} \rangle$ is expressed by a nonzero expectation value of $\langle \hat{\bm{l}} \times \hat{\bm{\sigma}} \rangle$ but $\bm{l}\equiv \langle \hat{\bm{l}} \rangle=0$ and $\bm{\sigma}\equiv \langle \hat{\bm{\sigma}} \rangle=0$ as shown in Fig.~\ref{Fig:ETD}(b).
In other words, the atomic-scale ET dipole moment exists even without a local electric polarization.
One finds that $\braket{\hat{\bm{G}}}$ becomes nonzero when the vertical mirror symmetry is lost while keeping $\mathcal{P}$ and $\mathcal{T}$ symmetries, which corresponds to the time-reversal-even axial vector as Eq.~(\ref{eq:ETD_classical}).
The ET dipole is ubiquitous in the orbitally degenerate systems which consist of the $p$, $d$, and $f$ orbitals with the spin degree of freedom since the magnetic dipole operators ($\hat{\bm{l}}$ and $\hat{\bm{\sigma}}$) exist in the Hilbert space~\cite{Hayami_PhysRevB.98.165110}.
We note that the ET dipole in Eq.~(\ref{eq:ETD_quantum}) is independent from the other multipoles such as the electric quadrupole.

Since the natural conjugate field of $\bm{G}$ is the rotation field of the crystal $\bm{\nabla} \times \bm{u}$ ($\bm{u}$ represents a displacement vector field), $\bm{G}$ is not coupled directly to the electromagnetic fields.
Nevertheless, we show that $\bm{G}$ still affects the electromagnetic responses as a rotator.
As expected from the expression in Eq.~(\ref{eq:ETD_quantum}), nonzero $\bm{G}$ activates the effective spin-orbit coupling as proportional to $\bm{G}\cdot(\bm{l}\times \bm{\sigma})$. 
In such a situation, the external magnetic field $\bm{H}$ induces the transverse component of the magnetization $\bm{m} \perp \bm{H}$ as shown in Fig.~\ref{Fig:ETD}(c)~\cite{cheong2021permutable}, since both $\bm{l}$ and $\bm{\sigma}$ are conjugate to $\bm{H}$~\cite{comment_rot}.
This rotating field can be detected by the magnetic torque measurement.

\begin{figure}[t!]
\begin{center}
\includegraphics[width=1.0 \hsize]{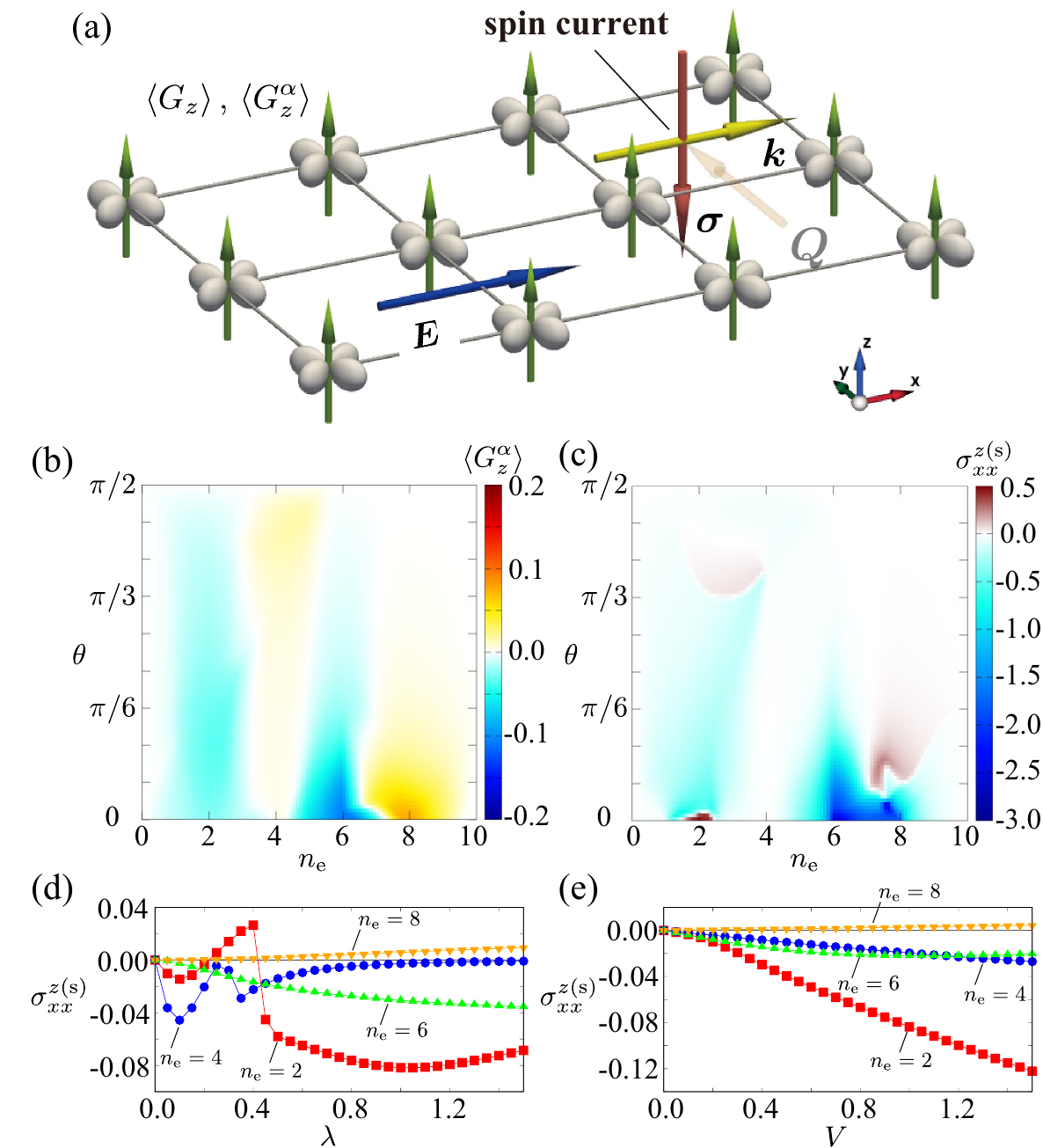}
\caption{
\label{Fig:SC}
(a) Schematic of spin current generation under the ET moment; the longitudinal spin current $J_{x}^{z{\rm (s)}} \propto Q_{y}(\bm{k}) = \left(\bm{k} \times \bm{\sigma}\right)_{y}$ is obtained along the applied electric field $\bm{E} \parallel \hat{\bm{x}}$.
Contour plots of (b) $\langle G^\alpha_z \rangle$ and (c) $\sigma^{z {\rm (s)}}_{xx}$ in the plane of $n_{\rm e}$ and $\theta$ at $V=0.7$.
(d) $\lambda$ and (e) $V$ dependences of $\sigma^{z {\rm (s)}}_{xx}$ for different $n_{\rm e}$ at (d) $V=0.7$ and (e) $\lambda=4$ for $\theta=\pi/3$.
}
\end{center}
\end{figure}

Similarly, nonzero $\braket{\hat{\bm{G}}}$ works as a rotator to the other fields, such as the electric field $\bm{E}$ and electric current $\bm{J}$.
To demonstrate it phenomenologically, we express the ET dipole according to the symmetry as
\begin{align}
\label{eq:ETD_quantum2}
\braket{\hat{\bm{G}}} \quad \leftrightarrow
\quad \bm{X}(\bm{r})\times \bm{X}(\bm{k}),
\end{align}
where $\bm{X}(\bm{r})$ and $\bm{X}(\bm{k})$ represent either of two dipoles ($\bm{X}=\bm{Q}, \bm{T}$) in real and momentum spaces, respectively. 
The expressions of $\bm{X}(\bm{r})$ and $\bm{X}(\bm{k})$ for $\bm{X}=\bm{Q}, \bm{M}, \bm{T}$ are summarized in Table~\ref{tab_irrep} ($\bm{k}$ is the wave vector and we take $\hbar=1$).
Equation~(\ref{eq:ETD_quantum2}) indicates that nonzero $\braket{\hat{\bm{G}}}$ turns on an effective transverse coupling between $\bm{X}(\bm{r})$ and $\bm{X}(\bm{k})$, which results in a rotated response of physical quantities against the applied $\bm{E}$ and $\bm{J}$ as similar to $\bm{H}$.
By noting that $\bm{E}$ ($\bm{J}$) is the conjugate field to $\bm{Q}(\bm{r})$ [$\bm{T}(\bm{k})$], one expects that a nonzero polar vector quantity with the same symmetry as $\bm{Q}$ ($\bm{T}$) is induced along the perpendicular direction to $\bm{E}$ ($\bm{J}$), as shown in Fig.~\ref{Fig:ETD}(c)~\cite{cheong2021permutable,comment_rot}.
For example, the electric polarization $P_y$ ($-P_x$) is induced under $E_x$ ($E_y$) in the presence of $\braket{\hat{G}_z}$~\cite{Nasu_PhysRevB.105.245125}.
In addition, other physical quantities with the same symmetry as $\bm{Q}$, such as the spin current and temperature gradient, are also induced by $\bm{E}$, as shown below in the case of the spin current [see Fig.~\ref{Fig:SC}(a)].

In a crystal with discrete symmetry, some components of the higher-rank ET octupoles $\hat{\bm{G}}^\alpha=(\hat{G}_x^\alpha,\hat{G}_y^\alpha,\hat{G}_z^\alpha)$ also contribute to the above electromagnetic responses, since they belong to the same irreducible representation as that of $\hat{\bm{G}}$. 
The ET octupole operator $\hat{G}^\alpha_z$ is obtained by replacing the magnetic dipole operator $(\hat{l}_x, \hat{l}_y)$ with the magnetic octupole operator $(\hat{M}_{3u}, \hat{M}_{3v})$ in the $z$ component of Eq.~(\ref{eq:ETD_quantum}) [Fig.~\ref{Fig:ETD}(d)]:
\begin{align}
\label{eq:ETD_quantum_octu}
\hat{G}_z^{\alpha}&=\hat{M}_{3u}(\bm{r}) \hat{\sigma}_y -\hat{M}_{3v}(\bm{r}) \hat{\sigma}_x \\
&
\leftrightarrow \quad[\bm{X}^{(3)}(\bm{r})\times \bm{X}(\bm{k})]_z ,
\end{align}
where $\bm{X}^{(3)}=(X_{3u}, X_{3v})$ for $\bm{X}=\bm{Q}, \bm{T}$ is represented by the linear combination of the spherical harmonics, $Y_{31}$ and $Y_{3-1}$~\cite{Hayami_PhysRevB.98.165110}.
As the octupole $(X_{3u}, X_{3v})$ belongs to the same irreducible representation as that of the dipole $(X_{x}, X_{y})$, the same discussion holds as the case of the ET dipole.
Among the 122 magnetic point groups, $\braket{\hat{G}_z}$ and $\braket{\hat{G}^\alpha_z}$ become nonzero for 43 magnetic point groups without the vertical mirror symmetry~\cite{comment_PG}.
Hereafter, we omit the hat symbol for operators for notational simplicity.

\begin{figure}[t!]
\begin{center}
\includegraphics[width=1.0 \hsize]{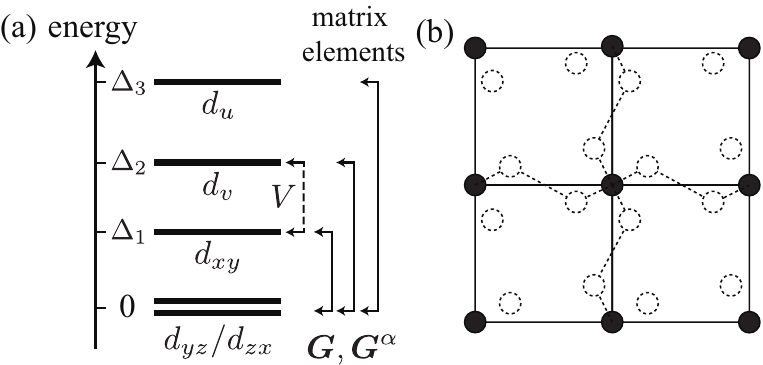}
\caption{
\label{Fig:model}
(a) CEF splitting and active ET multipoles within the five $d$ orbitals.
$V$ represents the hybridization between the $d_{v}$ and $d_{xy}$ orbitals.
(b) The crystal structure under the point group $C_{4{\rm h}}$ with $V\ne 0$.
The filled (dashed) circles denote the $d$-orbital (ligand) sites.
}
\end{center}
\end{figure}

{\it Lattice model.---}
To discuss the effect of the ET moment microscopically, we analyze a specific lattice model on the square lattice with five $d$ orbitals.
Let us start from the tight-binding model under the point group $D_{\rm 4h}$ preserving the vertical mirror symmetry, whose Hamiltonian is given by
\begin{align}
\label{eq:Ham}
\mathcal{H}&=\mathcal{H}_t + \mathcal{H}_{\rm SOC}\cos \theta +  \mathcal{H}_{\rm CEF} \sin \theta,\\
\label{eq:Hamt}
\mathcal{H}_t&= \sum_{ij \alpha\beta\sigma}(t_{\alpha\beta}^{ij}  c^{\dagger}_{i \alpha \sigma}c_{j \beta\sigma}+{\rm H.c.}),\\
\label{eq:HamSOC}
\mathcal{H}_{\rm SOC}&= \lambda \sum_i \bm{l}_i\cdot \bm{s}_i,  \\
\label{eq:HamCEF}
\mathcal{H}_{\rm CEF}&= \sum_{i\sigma} (
\Delta_1 c^{\dagger}_{i xy \sigma}c_{i xy \sigma}
+\Delta_2 c^{\dagger}_{i v \sigma}c_{i v \sigma}
+\Delta_3 c^{\dagger}_{i u \sigma}c_{i u \sigma}
),
\end{align}
where $c^\dagger_{i\alpha\sigma}$ ($c_{i\alpha \sigma}$) is the creation (annihilation) operator of a conduction electron with site $i$, orbital $(u=3z^2-r^2, v=x^2-y^2, yz, zx, xy)$ for five $d$ orbitals ($d_{u}, d_{v}, d_{yz}, d_{zx}, d_{xy}$), and spin $\sigma$.
$\mathcal{H}_t$ in Eq.~(\ref{eq:Hamt}) represents the kinetic energy of electrons.
We consider the nearest-neighbor hopping satisfying the Slater-Koster parameters: $t_{dd\sigma}=-1$ and $t_{dd\pi}=0.5$ ($t_{dd\sigma}$ is taken as the energy unit).
$\mathcal{H}_{\rm SOC}$ in Eq.~(\ref{eq:HamSOC}) stands for the atomic spin-orbit coupling with $\lambda=4$ where $\bm{l}_i$ and $\bm{s}_i$ represent the orbital- and spin-angular momentum operators.
$\mathcal{H}_{\rm CEF}$ in Eq.~(\ref{eq:HamCEF}) denotes the crystalline electric field (CEF) under $D_{\rm 4h}$; we take $\Delta_1=0.5$, $\Delta_2=6.5$, and $\Delta_3=10.5$, as shown in Fig.~\ref{Fig:model}(a).
In addition, we introduce the parameter $\theta$ to discuss the SOC contribution by adjusting the weight of $\mathcal{H}_{\rm SOC}$ and $\mathcal{H}_{\rm CEF}$, where $\theta=0$ ($\theta=\pi/2$) represents the strong (negligible) spin-orbit coupling regime.
Although we chose the specific values of model parameters ($t_{dd\sigma}, t_{dd\pi}, \lambda$, $\Delta_1$, $\Delta_2$, $\Delta_3$) so that the energy scale of $\mathcal{H}_t$, $\mathcal{H}_{\rm SOC}$, and $\mathcal{H}_{\rm CEF}$ are comparable with $4d$- and $5d$-electron systems in mind, the qualitative tendency is not altered for a different set of parameters.

The ET multipoles $(G_z, G_z^\alpha)$ belonging to the ${\rm A}^+_{2g}$ representation ($+$ in the superscript means the time-reversal even) are the candidate microscopic variables of the ``order parameter'' characterizing the nonzero ferro-axial moment when the symmetry is lowered to $C_{\rm 4h}$ as shown below.
As the Hamiltonian in Eq.~(\ref{eq:Hamt}) is a $10 \times 10$ matrix spanned by five $d$ orbitals with two spin components, the system has totally $10\times 10=100$ independent electronic multipoles.
Among them, there are only four multipoles in the ${\rm A}^+_{2g}$ representation;
one ET dipole $G_z$, and one ET octupole $G_z^\alpha$, and two electric hexadecapole $Q^\alpha_{4z}, Q'^{\alpha}_{4z}$.
$Q^\alpha_{4z} \propto xy (x^2-y^2)$ is defined in the spinless atomic basis, while the other three multipoles are defined in the spinful atomic basis, e.g., $G_z$ and $G_z^\alpha$ are represented by Eqs.~(\ref{eq:ETD_quantum}) and (\ref{eq:ETD_quantum_octu}), respectively~\cite{SM_ETD}.

The nonzero ET moment appears when the vertical mirror symmetry is lost.
We here consider the situation where the existence of the ligand ions break the vertical mirror symmetry, as shown in the dashed circles in Fig.~\ref{Fig:model}(b).
Then, the lattice symmetry is lowered from $D_{4\rm h}$ to $C_{4\rm h}$, and accordingly, there is additional CEF term which hybridizes $d_{v}$ and $d_{xy}$ orbitals:
\begin{align}
\label{eq:HamV}
\mathcal{H}_{V}&= V \sum_{i \sigma}(c^{\dagger}_{i xy \sigma}c_{i v \sigma}+c^{\dagger}_{i v \sigma}c_{i xy \sigma}),
\end{align}
which leads to nonzero $\braket{Q^\alpha_{4z}}$.
In the following, we analyze the total Hamiltonian, $\mathcal{\tilde{H}}=\mathcal{H}+\mathcal{H}_V$.

When $V \neq 0$, we obtain the nonzero ET moment, $\langle G_z^\alpha \rangle \neq 0$ but $\langle G_z \rangle = 0$~\cite{comment_atomicG}.
We show the color plot of $\langle G_z^\alpha \rangle$ while changing $\theta$ and the electron filling $n_{\rm e}$ ($n_{\rm e}=10$ represents the full-filling case) at $V=0.7$ in Fig.~\ref{Fig:SC}(b).
Although $V$ mixes only $d_{v}$ and $d_{xy}$ orbitals, nonzero $\langle G_z^\alpha \rangle$ is obtained in all the regions for $\theta$ and $n_{\rm e}$.
This is because the operator $G^{\alpha}_z$ consists of the off-diagonal elements connecting between all the $d$ orbitals, as shown in Fig.~\ref{Fig:model}(a).
The essential model parameters to give nonzero $\langle G_z^\alpha \rangle$ are given by $V \lambda t_{dd\sigma}t_{dd\pi}$~\cite{Hayami_PhysRevB.102.144441,oiwa2021systematic,SM_ETD}; the spin-orbit coupling $\lambda$ and the hoppings $t_{dd\sigma}$ and $t_{dd\pi}$ play an important role in addition to $V$.
This indicates that nonzero independent $\langle G_z^\alpha \rangle$ is obtained only when the system includes at least four $d$-orbitals ($d_{v}, d_{yz}, d_{zx}, d_{xy}$) and both $\mathcal{H}_{\rm SOC}$ and $\mathcal{H}_V$ become relevant.
In other words, the situation, where the cubic/tetragonal CEF is comparable to the atomic spin-orbit coupling, such as $4d$ and $5d$ electrons, is preferred to activate the ET moment.

\begin{figure}[t!]
\begin{center}
\includegraphics[width=1.0 \hsize]{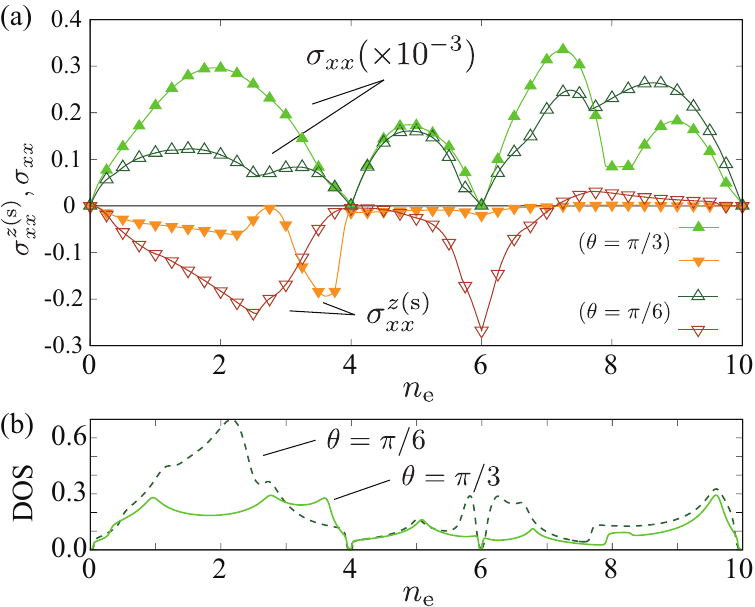}
\caption{
\label{Fig:SC_filling}
(a) $n_{\rm e}$ dependences of $\sigma^{z {\rm (s)}}_{xx}$ and $\sigma_{xx}$ for $\theta=\pi/3$ and $\theta=\pi/6$ at $V=0.7$.
(b) Density of states (DOS) corresponding to (a).
}
\end{center}
\end{figure}

{\it Spin current generation.---}
When the nonzero ET moment exists, we expect the rotated electromagnetic responses as discussed above.
Among them, we find a new type of spin-current generation through the coupling $\bm{Q}(\bm{r}) \times \bm{Q}(\bm{k})$.
We calculate the spin-conductivity tensor defined by $J^{\eta{\rm (s)}}_{\nu} = \sum_{\mu}\sigma^{\eta {\rm (s)}}_{\mu\nu} E_{\mu}$, where $J^{\eta{\rm (s)}}_{\nu}= J_{\nu} \sigma_\eta$ is the spin-current operator, by using the Kubo formula with the scattering rate $\tau^{-1}=10^{-2}$ and the temperature $T=10^{-3}$.
Hereafter, we set $-e/4\pi$ as the unit of $\sigma^{\eta {\rm (s)}}_{\mu\nu}$ ($e$ is the elementary charge).
The summation of the momentum $\bm{k}$ is taken over $3600^2$ grid points in the first Brillouin zone.
In the $C_{\rm 4h}$ symmetry in general, we expect nonzero $\sigma^{\eta {\rm (s)}}_{xy}=-\sigma^{\eta {\rm (s)}}_{yx}$ and $\sigma^{\eta {\rm (s)}}_{xx}=\sigma^{\eta {\rm (s)}}_{yy}$ for $\eta=z$.
The former represents a conventional spin Hall effect~\cite{murakami2003dissipationless,Sinova_PhysRevLett.92.126603}, while the latter represents a new type of spin current characteristic of the nonzero ET moment. 
The latter has been recently studied from the symmetry viewpoint~\cite{Seemann_PhysRevB.92.155138,Akzyanov_PhysRevB.99.045436,roy2021unconventional}. 
Since $\sigma^{\eta {\rm (s)}}_{\mu\nu}$ corresponds to the rank-3 tensor, the rank 0-3 multipoles are important, such as the ET octupole.
In other words, the latter $\sigma^{z {\rm (s)}}_{xx}$ is driven by the coupling $[\bm{Q}(\bm{r}) \times \bm{Q}(\bm{k})]_z \sim (E_x k_x + E_y k_y )\sigma_z $, where $\bm{Q}(\bm{r})$ [$\bm{Q}(\bm{k})$] corresponds to the input (output) field [Fig.~\ref{Fig:SC}(a)].
Note that $\sigma_{xy}^{z({\rm s})}=0$ accidentally in the present model; the consideration of the second-neighbor hopping in Eq.~(\ref{eq:Hamt}) leads to nonzero $\sigma_{xy}^{z({\rm s})}$.

Figure~\ref{Fig:SC}(c) shows $\sigma^{z {\rm (s)}}_{xx}$ in the plane of $n_{\rm e}$ and $\theta$, indicating that nonzero $\sigma^{z {\rm (s)}}_{xx}$ is obtained for nonzero $\langle G_z^\alpha \rangle$~\cite{SM_ETD}.
Indeed, the essential model parameters to obtain finite $\sigma^{z {\rm (s)}}_{xx}$ are represented by $V \lambda^2 t_{dd\pi}$, which are similar to those for $\langle G_z^\alpha \rangle$~\cite{comment_Q4z,SM_ETD}.
We also show $\lambda$ and $V$ dependences of $\sigma^{z {\rm (s)}}_{xx}$ at $\theta=\pi/3$ for several $n_{\rm e}$ in Figs.~\ref{Fig:SC}(d) and \ref{Fig:SC}(e), where $\sigma^{z {\rm (s)}}_{xx}=0$ for $\lambda=0$ or $V=0$.
The non-monotonic behavior of $\sigma^{z {\rm (s)}}_{xx}$ in Fig.~\ref{Fig:SC}(d) indicates that large $\lambda$ is not necessary for large $\sigma^{z {\rm (s)}}_{xx}$.

Finally, we discuss an intriguing feature of the spin current under the nonzero ferro-axial moment.
Figure~\ref{Fig:SC_filling}(a) represents $n_{\rm e}$ dependence of $\sigma^{z {\rm (s)}}_{xx}$ at $V=0.7$ for $\theta=\pi/3$ and $\theta=\pi/6$.
We also show the longitudinal electric conductivity $\sigma_{xx}$ calculated by the same condition as $\sigma^{z {\rm (s)}}_{xx}$.
Notably, $\sigma^{z {\rm (s)}}_{xx}$ takes nonzero values even in the insulating case at $n_{\rm e}=4, 6$, where the density of states in Fig.~\ref{Fig:SC_filling}(b) vanish and $\sigma_{xx} \simeq 0$.
This is because $\sigma^{z {\rm (s)}}_{xx}$ is driven by the intrinsic interband process like the conventional spin Hall effect, which is in marked contrast to the symmetric spin current obtained in the antiferromagnetic metals~\cite{kimata2019magnetic,naka2019spin,Mook_PhysRevResearch.2.023065,hayami2022spinconductivity}.
Thus, one can generate ``the pure longitudinal spin current" along the applied electric field in the insulators with the ferro-axial moment.
Moreover, as $\sigma^{z {\rm (s)}}_{xx}$ is almost independent of $\tau^{-1}$, it is expected that the efficient spin-current generation is achieved in relatively dirty metals.

{\it Conclusion.---}
We theoretically investigated the ET-related physics on the basis of symmetry and microscopic analyses.
We showed the nonzero expectation value of the atomic-scale ET multipoles as an indicator of the ferro-axial ``ordering''.
We presented the microscopic conditions of the active ET moment: the multi-orbital systems including the spin-orbit coupling $\lambda$ and the hybridization $V$ breaking the vertical mirror symmetry.
We also discussed that the nonzero ET moment brings about rich physics in contrast to the conventional wisdom.
The ET moment acts as a nanometric rotator against external electromagnetic fields.
In particular, we demonstrated an intrinsic generation of the longitudinal spin current by the applied electric field.
Our conclusions are widely applicable to the systems belonging to 43 magnetic point groups with the ET moment.
The candidate materials to exhibit present ET physics are Ca(BH$_4$)$_2$, PdS, TaPO$_5$, VPO$_5$, and NbTe$_2$ that belong to the point group $C_{\rm 4h}$.
In addition, CaMn$_7$O$_{12}$~\cite{Johnson_PhysRevLett.108.067201}, RbFe(MoO$_4$)$_2$~\cite{jin2020observation}, NiTiO$_3$~\cite{hayashida2020visualization}, and Ca$_5$Ir$_3$O$_{12}$~\cite{hanate2021first} are also the candidates since they have been identified as the spontaneous ferro-axial orders having nonzero ET moment.

\begin{acknowledgments}
The authors would like to thank J. Nasu for fruitful discussions.
This research was supported by JSPS KAKENHI Grants Numbers JP19K03752, JP19H01834, JP21H01037, JP22H04468, JP22H00101, JP22H01183, and by JST PREST (JPMJPR20L8).
Parts of the numerical calculations were performed in the supercomputing systems in ISSP, the University of Tokyo, and in the MAterial science Supercomputing system for Advanced MUlti-scale simulations towards NExt-generation-Institute for Materials Research (MASAMUNE-IMR) of the Center for Computational Materials Science, Institute for Materials Research, Tohoku University.
\end{acknowledgments}

\bibliographystyle{JPSJ}
\bibliography{ref}

\begin{thebibliography}{10}

\bibitem{Spaldin_0953-8984-20-43-434203}
N.~A. Spaldin, M.~Fiebig, and M.~Mostovoy, J. Phys.: Condens. Matter {\bfseries
  20},  434203 (2008).

\bibitem{KhomskiiPhysics.2.20}
D.~Khomskii, Physics {\bfseries 2},  20 (2009).

\bibitem{kopaev2009toroidal}
Y.~V. Kopaev, Physics-Uspekhi {\bfseries 52},  1111 (2009).

\bibitem{Hayami_PhysRevB.90.024432}
S.~Hayami, H.~Kusunose, and Y.~Motome, Phys. Rev. B {\bfseries 90},  024432
  (2014).

\bibitem{aizu1966possible}
K.~Aizu, Phys. Rev. {\bfseries 146},  423 (1966).

\bibitem{aizu1969possible}
K.~Aizu, J. Phys. Soc. Jpn. {\bfseries 27},  387 (1969).

\bibitem{aizu1970possible}
K.~Aizu, Phs. Rev. B {\bfseries 2},  754 (1970).

\bibitem{Schmid_doi:10.1080/00150199408245120}
H.~Schmid, Ferroelectrics {\bfseries 162},  317 (1994).

\bibitem{wadhawan2000introduction}
V.~Wadhawan: {\em Introduction to ferroic materials} (CRC press, 2000).

\bibitem{schmid2008some}
H.~Schmid, J. Phys. Condens. Matter {\bfseries 20},  434201 (2008).

\bibitem{litvin2013magnetic}
D.~B. Litvin: {\em Magnetic Group Tables} (International Union of
  Crystallography, 2013).

\bibitem{Hayami_PhysRevB.98.165110}
S.~Hayami, M.~Yatsushiro, Y.~Yanagi, and H.~Kusunose, Phys. Rev. B {\bfseries
  98},  165110 (2018).

\bibitem{cheong2018broken}
S.-W. Cheong, D.~Talbayev, V.~Kiryukhin, and A.~Saxena, npj Quantum Mater.
  {\bfseries 3},  19 (2018).

\bibitem{Watanabe_PhysRevB.98.245129}
H.~Watanabe and Y.~Yanase, Phys. Rev. B {\bfseries 98},  245129 (2018).

\bibitem{Yatsushiro_PhysRevB.104.054412}
M.~Yatsushiro, H.~Kusunose, and S.~Hayami, Phys. Rev. B {\bfseries 104},
  054412 (2021).

\bibitem{parkin1983magnetic}
S.~Parkin, E.~Marseglia, and P.~Brown, J. Phys. C: Solid State Phys. {\bfseries
  16},  2765 (1983).

\bibitem{nakatsuji2015large}
S.~Nakatsuji, N.~Kiyohara, and T.~Higo, Nature {\bfseries 527},  212 (2015).

\bibitem{Suzuki_PhysRevB.95.094406}
M.-T. Suzuki, T.~Koretsune, M.~Ochi, and R.~Arita, Phys. Rev. B {\bfseries 95},
   094406 (2017).

\bibitem{ghimire2018large}
N.~J. Ghimire, A.~Botana, J.~Jiang, J.~Zhang, Y.-S. Chen, and J.~Mitchell, Nat.
  Commun. {\bfseries 9},  3280 (2018).

\bibitem{Tenasini_PhysRevResearch.2.023051}
G.~Tenasini, E.~Martino, N.~Ubrig, N.~J. Ghimire, H.~Berger, O.~Zaharko, F.~Wu,
  J.~F. Mitchell, I.~Martin, L.~Forr\'o, and A.~F. Morpurgo, Phys. Rev.
  Research {\bfseries 2},  023051 (2020).

\bibitem{Chen_PhysRevB.101.104418}
H.~Chen, T.-C. Wang, D.~Xiao, G.-Y. Guo, Q.~Niu, and A.~H. MacDonald, Phys.
  Rev. B {\bfseries 101},  104418 (2020).

\bibitem{vsmejkal2020crystal}
L.~{\v{S}}mejkal, R.~Gonz{\'a}lez-Hern{\'a}ndez, T.~Jungwirth, and J.~Sinova,
  Sci. Adv. {\bfseries 6},  eaaz8809 (2020).

\bibitem{Akiba_PhysRevResearch.2.043090}
K.~Akiba, K.~Iwamoto, T.~Sato, S.~Araki, and T.~C. Kobayashi, Phys. Rev.
  Research {\bfseries 2},  043090 (2020).

\bibitem{Hayami_PhysRevB.103.L180407}
S.~Hayami and H.~Kusunose, Phys. Rev. B {\bfseries 103},  L180407 (2021).

\bibitem{Shao_PhysRevApplied.15.024057}
D.-F. Shao, J.~Ding, G.~Gurung, S.-H. Zhang, and E.~Y. Tsymbal, Phys. Rev.
  Appl. {\bfseries 15},  024057 (2021).

\bibitem{Mangelsen_PhysRevB.103.184408}
S.~Mangelsen, P.~Zimmer, C.~N\"ather, S.~Mankovsky, S.~Polesya, H.~Ebert, and
  W.~Bensch, Phys. Rev. B {\bfseries 103},  184408 (2021).

\bibitem{park2021first}
H.~Park, O.~Heinonen, and I.~Martin, arXiv:2110.03029 ,  (2021).

\bibitem{dubovik1986axial}
V.~Dubovik, L.~Tosunyan, and V.~Tugushev, Zh. Eksp. Teor. Fiz {\bfseries 90},
  590 (1986).

\bibitem{dubovik1990toroid}
V.~Dubovik and V.~Tugushev, Phys. Rep. {\bfseries 187},  145 (1990).

\bibitem{Hlinka_PhysRevLett.113.165502}
J.~Hlinka, Phys. Rev. Lett. {\bfseries 113},  165502 (2014).

\bibitem{dubovik1975multipole}
V.~Dubovik and A.~Cheshkov, Sov. J. Part. Nucl {\bfseries 5},  318 (1975).

\bibitem{Johnson_PhysRevLett.108.067201}
R.~D. Johnson, L.~C. Chapon, D.~D. Khalyavin, P.~Manuel, P.~G. Radaelli, and
  C.~Martin, Phys. Rev. Lett. {\bfseries 108},  067201 (2012).

\bibitem{yadav2016observation}
A.~Yadav, C.~Nelson, S.~Hsu, Z.~Hong, J.~Clarkson, C.~Schlep{\"u}tz,
  A.~Damodaran, P.~Shafer, E.~Arenholz, L.~Dedon, D.~Chen, A.~Vishwanath, A.~M.
  Minor, L.~Q. Chen, J.~F. Scott, L.~W. Martin, and R.~Ramesh, Nature
  {\bfseries 530},  198 (2016).

\bibitem{jin2020observation}
W.~Jin, E.~Drueke, S.~Li, A.~Admasu, R.~Owen, M.~Day, K.~Sun, S.-W. Cheong, and
  L.~Zhao, Nat. Phys. {\bfseries 16},  42 (2020).

\bibitem{hayashida2020visualization}
T.~Hayashida, Y.~Uemura, K.~Kimura, S.~Matsuoka, D.~Morikawa, S.~Hirose,
  K.~Tsuda, T.~Hasegawa, and T.~Kimura, Nat. Commun. {\bfseries 11},  4582
  (2020).

\bibitem{hanate2021first}
H.~Hanate, T.~Hasegawa, S.~Hayami, S.~Tsutsui, S.~Kawano, and K.~Matsuhira, J.
  Phys. Soc. Jpn. {\bfseries 90},  063702 (2021).

\bibitem{hayami2018microscopic}
S.~Hayami and H.~Kusunose, J. Phys. Soc. Jpn. {\bfseries 87},  033709 (2018).

\bibitem{kusunose2020complete}
H.~Kusunose, R.~Oiwa, and S.~Hayami, J. Phys. Soc. Jpn. {\bfseries 89},  104704
  (2020).

\bibitem{Hlinka_PhysRevLett.116.177602}
J.~Hlinka, J.~Privratska, P.~Ondrejkovic, and V.~Janovec, Phys. Rev. Lett.
  {\bfseries 116},  177602 (2016).

\bibitem{comment_PG}
The 43 magnetic point groups with the active ET moments $\braket{\bm{G}},
  \braket{\bm{G}^\alpha}$ are $4/m1'$, $41'$, $\bar{4}1'$, $6/m1'$, $61'$,
  $\bar{6}1'$, $\bar{3}1'$, $31'$, $2/m1'$, $21'$, $m1'$, $\bar{1}1'$, $11'$,
  $4/m$, $4$, $\bar{4}$, $6/m$, $6$, $\bar{6}$, $\bar{3}$, $3$, $2/m$, $2$,
  $m$, $\bar{1}$, $1$, $4'/m'$, $4/m'$, $4'/m$, $4'$, $\bar{4}'$, $6'/m'$,
  $6/m'$, $6'/m$, $6'$, $\bar{6}'$, $\bar{3}'$, $2'/m'$, $2/m'$, $2'/m$, $2'$,
  $m'$, and $\bar{1}'$.

\bibitem{cheong2021permutable}
S.-W. Cheong, S.~Lim, K.~Du, and F.-T. Huang, npj Quantum Mater. {\bfseries 6},
   58 (2021).

\bibitem{comment_rot}
Since the linear dielectric (magnetic susceptibility) tensor is a symmetric
  tensor, the third-order electric polarization (magnetization) is expected to
  be nonzero.

\bibitem{Nasu_PhysRevB.105.245125}
J.~Nasu and S.~Hayami, Phys. Rev. B {\bfseries 105},  245125 (2022).

\bibitem{SM_ETD}
See Supplemental Materials at (xx) for details about the electronic multipoles
  in the five $d$-orbital system and the essential model parameters of the ET
  moment and the spin-conductivity tensor.

\bibitem{comment_atomicG}
The disappearance of $\langle G_z \rangle$ is presumably owing to the symmetry
  of the system and/or the lack of the microscopic ingredients in the
  Hamiltonian. For example, nonzero $\langle G_z \rangle$ occurs in the
  lower-symmetry system under the vortex-type spin
  configuration~\cite{hayami_vortex}.

\bibitem{Hayami_PhysRevB.102.144441}
S.~Hayami, Y.~Yanagi, and H.~Kusunose, Phys. Rev. B {\bfseries 102},  144441
  (2020).

\bibitem{oiwa2021systematic}
R.~Oiwa and H.~Kusunose, J. Phys. Soc. Jpn. {\bfseries 91},  014701 (2022).

\bibitem{murakami2003dissipationless}
S.~Murakami, N.~Nagaosa, and S.-C. Zhang, Science {\bfseries 301},  1348
  (2003).

\bibitem{Sinova_PhysRevLett.92.126603}
J.~Sinova, D.~Culcer, Q.~Niu, N.~A. Sinitsyn, T.~Jungwirth, and A.~H.
  MacDonald, Phys. Rev. Lett. {\bfseries 92},  126603 (2004).

\bibitem{Seemann_PhysRevB.92.155138}
M.~Seemann, D.~K\"odderitzsch, S.~Wimmer, and H.~Ebert, Phys. Rev. B {\bfseries
  92},  155138 (2015).

\bibitem{Akzyanov_PhysRevB.99.045436}
R.~S. Akzyanov and A.~L. Rakhmanov, Phys. Rev. B {\bfseries 99},  045436
  (2019).

\bibitem{roy2021unconventional}
A.~Roy, M.~H.~D. Guimar\~aes, and J.~S\l{}awi\ifmmode~\acute{n}\else
  \'{n}\fi{}ska, Phys. Rev. Materials {\bfseries 6},  045004 (2022).

\bibitem{comment_Q4z}
$Q'^{\alpha}_{4z}$ also contributes to $\sigma^{z {\rm (s)}}_{xx}$ through the
  effective coupling to the ET octupole, whose essential model parameter is
  given by $V \lambda$.

\bibitem{kimata2019magnetic}
M.~Kimata, H.~Chen, K.~Kondou, S.~Sugimoto, P.~K. Muduli, M.~Ikhlas, Y.~Omori,
  T.~Tomita, A.~H. MacDonald, S.~Nakatsuji, and Y.~Otani, Nature {\bfseries
  565},  627 (2019).

\bibitem{naka2019spin}
M.~Naka, S.~Hayami, H.~Kusunose, Y.~Yanagi, Y.~Motome, and H.~Seo, Nat. Commun.
  {\bfseries 10},  4305 (2019).

\bibitem{Mook_PhysRevResearch.2.023065}
A.~Mook, R.~R. Neumann, A.~Johansson, J.~Henk, and I.~Mertig, Phys. Rev.
  Research {\bfseries 2},  023065 (2020).

\bibitem{hayami2022spinconductivity}
S.~Hayami and M.~Yatsushiro, J. Phys. Soc. Jpn. {\bfseries 91},  063702 (2022).

\bibitem{hayami_vortex}
S.~Hayami, Phys. Rev. B {\bfseries 106},  144402 (2022).

\end{thebibliography}

\end{document}